%% file: main.tex
\newif\ifdouble
\newif\ifsingle
\newif\ifchange
\doubletrue

\ifdouble
  \documentclass[sigconf, usenames, dvipsnames]{acmart}
\else
  \documentclass[sigconf,manuscript]{acmart}
  \singletrue
\fi

\usepackage{dblfloatfix} 
\usepackage{booktabs} 
\usepackage{arydshln}
\usepackage{graphicx,subcaption}
\newcommand{\added}[1]{{\color{purple}{#1}}}
\renewcommand{\added}[1]{#1}

\newcommand{\removed}[1]{\iffalse #1 \fi}

\newcommand{\halfwidth}{1\linewidth}

\newcommand{\subsub}[1]{\noindent \textit{\textbf{#1:}}}

\AtBeginDocument{%
  \providecommand\BibTeX{{%
    \normalfont B\kern-0.5em{\scshape i\kern-0.25em b}\kern-0.8em\TeX}}}
    
\copyrightyear{2023}
\acmYear{2023}
\setcopyright{acmlicensed}\acmConference[UIST '23]{The 36th Annual ACM Symposium on User Interface Software and Technology}{October 29-November 1, 2023}{San Francisco, CA, USA}
\acmBooktitle{The 36th Annual ACM Symposium on User Interface Software and Technology (UIST '23), October 29-November 1, 2023, San Francisco, CA, USA}
\acmPrice{15.00}
\acmDOI{10.1145/3586183.3606716}
\acmISBN{979-8-4007-0132-0/23/10}

\newcommand{\system}{RealityCanvas}

\begin{document}
\pagenumbering{arabic}
\pagestyle{plain}
\title{\system{}: Augmented Reality Sketching for Embedded and Responsive Scribble Animation Effects}



\author{Zhijie Xia}
\affiliation{%
  \institution{University of Calgary}
  \city{Calgary}
  \country{Canada}}
\email{zhijie.xia@ucalgary.ca}
\authornote{Both authors contributed equally to the paper}

\author{Kyzyl Monteiro}
\affiliation{%
  \institution{IIIT-Delhi}
  \city{Delhi}
  \country{
  India}}
\affiliation{%
  \institution{University of Calgary}
  \city{Calgary}
  \country{
  Canada}}  
\email{kyzyl17296@iiitd.ac.in}
\authornotemark[1]

\author{Kevin Van}
\affiliation{%
  \institution{University of Calgary}
  \city{Calgary}
  \country{Canada}}
\email{kevin.van@ucalgary.ca}

\author{Ryo Suzuki}
\affiliation{%
  \institution{University of Calgary}
  \city{Calgary}
  \country{Canada}}
\email{ryo.suzuki@ucalgary.ca}

\renewcommand{\shortauthors}{Xia and Monteiro, et al.}
\input{0-abstract}

\begin{teaserfigure}
\includegraphics[width=\textwidth]{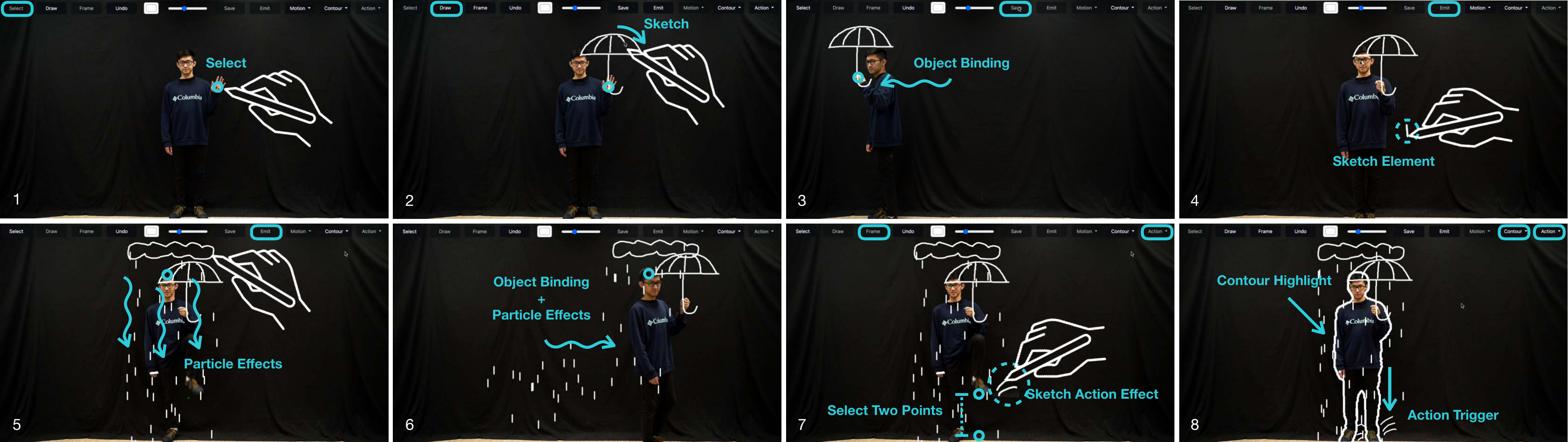}
\caption{\system{} workflow (from top left to bottom right): 1) select a hand as a tracking point, 2) sketch an umbrella bound to the hand, 3) the umbrella moves when the hand moves, 4) sketch a raindrop, 5) draw a cloud as an emitter line to show particle effects, 6) the cloud also moves with object binding, 7) select the ground and right foot then sketch a water splash, 8) show the water splash and contour highlight based on the stomp action.}
\label{fig:teaser}
\end{teaserfigure}

\maketitle

\input{1-introduction}

\input{2-related-work}
\input{3-design-space}
\input{4-implementation}

\input{5-application}
\input{6-user-study}

\input{7-future-work}

\input{8-conclusion}

\ifdouble
  \balance
\fi
\bibliographystyle{ACM-Reference-Format}
\bibliography{references}

\end{document}

%% file: 0-abstract.tex
\begin{abstract}
We introduce RealityCanvas, a mobile AR sketching tool that can easily augment real-world physical motion with responsive hand-drawn animation. Recent research in AR sketching tools has enabled users to not only embed static drawings into the real world but also dynamically animate them with physical motion. However, existing tools often lack the flexibility and expressiveness of possible animations, as they primarily support simple line-based geometry. To address this limitation, we explore both \textit{expressive} and \textit{improvisational} AR sketched animation by introducing a set of responsive scribble animation techniques that can be directly embedded through sketching interactions: 1) object binding, 2) flip-book animation, 3) action trigger, 4) particle effects, 5) motion trajectory, and 6) contour highlight. These six animation effects were derived from the analysis of 172 existing video-edited scribble animations. We showcase these techniques through various applications, such as video creation, augmented education, storytelling, and AR prototyping. The results of our user study and expert interviews confirm that our tool can lower the barrier to creating AR-based sketched animation, while allowing creative, expressive, and improvisational AR sketching experiences. 
\end{abstract}

\ccsdesc[500]{Human-centered computing~Mixed / augmented reality}

\keywords{Sketching Interfaces; Scribble Animation; Augmented Reality; Mixed Reality; Real-Time Authoring}

%% file: 1-introduction.tex
\section{Introduction}
In recent years, many augmented reality (AR) sketching tools~\cite{arora2018symbiosissketch, gasques2019pintar, suzuki2020realitysketch, leiva2020pronto} have been explored in human-computer interaction (HCI) research, thanks to the proliferation of augmented and mixed reality (AR/MR) devices.
In contrast to traditional screen-based sketching interfaces such as \textit{SketchPad} ~\cite{sutherland1964sketchpad}, \textit{Pegasus}~\cite{igarashi1998pegasus}, \textit{Draco}~\cite{kazi2014draco}, and many others~\cite{bae2008ilovesketch, gross1996ambiguous, landay1995interactive}, \textit{AR sketching} enables the user to sketch and annotate directly onto a live or recorded real-world scene, opening up untapped opportunities for many applications, such as 3D design and prototyping (e.g., \textit{SymbiosisSketch}~\cite{arora2018symbiosissketch}, \textit{DesignAR}~\cite{reipschlager2019designar}), education (e.g., \textit{Augmented Body}~\cite{ferdous2019s}, \textit{PaperTrail}~\cite{rajaram2022paper}), collaboration (e.g., \textit{Vuforia Chalk}~\cite{vuforia-chalk}, \textit{PintAR}~\cite{gasques2019pintar}), and entertainment (e.g., \textit{Just a Line}~\cite{just-a-line}, \textit{DoodleLens}~\cite{doodlelens}).

While many existing AR sketching tools focus on \textit{static sketching}, in which sketched elements float in the mid-air without being animated, more recent works have explored \textit{dynamic sketching}, in which the sketched elements can animate and respond to real-world interactions~\cite{suzuki2020realitysketch, saquib2022graphiti, saquib2019interactive, kaimoto2022sketched, perlin2018chalktalkvrar}.
Such dynamic AR sketching tools allow more interactive and engaging experiences for various applications, however, these existing tools either require both preparation and pre-defined configuration\removed{ and preparation}, hindering real-time and improvisational exploration (e.g., \textit{Interactive Body-Driven Graphics}~\cite{saquib2019interactive}, \textit{ChalkTalk AR}~\cite{perlin2018chalktalkvrar}\removed{, Pronto~\cite{leiva2020pronto}}) or they do not support freehand drawing except for simple line-based geometries, significantly limiting the flexibility and generalizability of possible animations (e.g., \textit{RealitySketch}~\cite{suzuki2020realitysketch}, \textit{Graphiti}~\cite{saquib2022graphiti}, \textit{Sketched Reality}~\cite{kaimoto2022sketched}).

To address these limitations, we introduce \system{}, a mobile AR sketching tool that allows the user to embed responsive sketched animation to a live or recorded real-world scene through \textit{\textbf{freehand}} and \textbf{\textit{real-time}} sketching interactions (Figure~\ref{fig:teaser}).
The goal of \system{} is to enable more \textit{expressive} yet \textit{improvisational} AR sketched animations that can respond to and interact with the real world (Figure~\ref{fig:related-work}).
\removed{To this end, we introduce \system{}, an AR sketching tool that allows the user to embed responsive sketched animation to a live or recorded real-world scene through \textit{\textbf{freehand}} and \textit{\textbf{improvisational}} sketching interactions (Figure~\ref{fig:teaser}).}
With both expressive and improvisational capabilities, \system{} allows users to spontaneously blend digital sketches and physical motion in a flexible manner, without the need for prior preparation or planning. This enables interactive exploration and experimentation in real time, unlike prior work~\cite{saquib2019interactive, perlin2018chalktalkvrar}.

To design our system, we first collect and analyze 172 existing video-edited scribble animation examples to inform us of the possible design space of expressive and comprehensive sketched animations which respond to real-world interactions (Figure~\ref{fig:design-space}).
Based on the analysis, we identify the following six most common scribble animation effects: 1) \textit{\textbf{object binding}}: dynamically move sketched elements based on the body or object movement, 2) \textit{\textbf{flip-book animation}}: creating an animation based on multiple sketched frames, 3) \textit{\textbf{action trigger}}: one-time animation effect based on the specific action, 4) \textit{\textbf{particle effects}}: spawning many sketched elements around the object, 5) \textit{\textbf{motion trajectory}}: showing the motion path of objects, and 6) \textit{\textbf{contour highlight}}: line-based animation around the contour of a body or object.

Our main contribution is a set of sketching interaction techniques that allow the user to quickly create all of these expressive scribble animations through direct manipulation with the following three step workflow: 
1) \textit{object tracking}: the user specifies a visual entity (e.g., a physical object, a skeletal joint) to track in a live or recorded real-world scene;
2) \textit{sketch elements}: the user adds sketched objects with freehand drawing;
3) \textit{animate sketched elements:} the sketches respond to the world based on the above animation effects.
\removed{With this, \system{} lets the user sketch responsive AR animation in \textit{real-time} and \textit{improvisational} ways, without pre-defined programs or post-production.} 
With these range of techniques through a simple workflow, \system{} lets the user sketch responsive AR animation in \textit{real-time} and in \textit{flexible} ways, without prior preparation or planning.

We demonstrate applications of \system{} \removed{such as}including social media video creation, augmented classroom education, and storytelling.
We evaluate our system through two user studies: 1) a usability study with twenty participants, and 2) expert reviews with seven professional video creators and theatre professionals.
The study results confirm that our tool lowers the barrier to creating AR-based sketched animation by enabling the end-user to quickly draw expressive animation in engaging and intuitive ways.

In summary, this paper contributes:
\begin{enumerate}
\item A taxonomy of existing scribble animation effects that informs us of the possible design space of expressive AR sketched animation.
\item A set of techniques that enable authoring of expressive freehand AR sketched animations through improvisational sketching interactions.
\item An implementation, applications, and the user evaluation of \system{}.
\end{enumerate}


%% file: 2-related-work.tex
\section{Related Work}

\subsection{Augmented Reality Sketching Tools}
With recent advances of AR technology, there is an increasing number of AR (or VR) sketching tools in both commercial products (e.g., \textit{Just a Line}~\cite{just-a-line}, \textit{TiltBrush}~\cite{tiltbrush}, \textit{Gravity Sketch}~\cite{gravity-sketch}, \textit{Vuforia Chalk AR}~\cite{vuforia-chalk}) as well as research prototypes (e.g., \textit{SymbiosisSketch}~\cite{arora2018symbiosissketch}, \textit{VRSketchIn}~\cite{drey2020vrsketchin}, \textit{PintAR}~\cite{gasques2019pintar}).
By leveraging the benefits of \textit{blending} virtual sketches into the physical world, these tools expand the real-world sketching experience for designing (e.g., \textit{SymbiosisSketch}~\cite{arora2018symbiosissketch}, \textit{Mobi3DSketch}~\cite{kwan2019mobi3dsketch}, \textit{DesignAR}~\cite{reipschlager2019designar}), AR prototyping (e.g., \textit{ProtoAR}~\cite{nebeling2018protoar}, \textit{360Proto}~\cite{nebeling2019360proto}), education (e.g., \textit{PaperTrail}~\cite{rajaram2022paper}, \textit{Augmented Body}~\cite{ferdous2019s}), collaboration (e.g., \textit{Vuforia Chalk}~\cite{vuforia-chalk}, \textit{PintAR}\cite{gasques2019pintar}), and entertainment (e.g., \textit{DoodleLens}~\cite{doodlelens}).

\removed{While most of these AR sketching tools only support \textit{static sketching}, in which sketched elements only stay in the mid-air without moving or animating,} 
To go beyond simple static AR sketching, more recent works have started exploring \textit{dynamic and responsive} sketched animation, in which the sketched elements can animate and respond to the corresponding real-world interaction.
For example, \textit{Interactive Body-driven Graphics}~\cite{saquib2019interactive} or \textit{ChalkTalk AR}~\cite{perlin2018chalktalkvrar} allows the creation of dynamic sketched animations for augmented storytelling.
\textit{RakugakiAR}~\cite{rakugakiar} allows the quick creation of animated sketched characters in AR.
Tools like \textit{Rapido}~\cite{leiva2021rapido} and \textit{Pronto}~\cite{leiva2020pronto} leverage expressive freehand sketched animations for video-based AR prototyping.
Moreover, \textit{RealitySketch}~\cite{suzuki2020realitysketch}, \textit{Reactile}~\cite{suzuki2018reactile}, \textit{Graphiti}~\cite{saquib2022graphiti}, and \textit{Sketched Reality}~\cite{kaimoto2022sketched} let the user bind sketched elements with a physical object so that the user can create responsive animations with real-time and improvisational sketching interactions.

However, these existing tools often require pre-defined configuration or preparation~\cite{saquib2019interactive, perlin2018chalktalkvrar, leiva2020pronto}.
For example, prior work~\cite{saquib2019interactive, perlin2018chalktalkvrar} distinguishes between a \textit{preparation phase} and \textit{interaction phase}, by requiring the user to prepare  assets~\cite{saquib2019interactive} or pre-programmed animations~\cite{perlin2018chalktalkvrar} prior to the performance, which hinders real-time and improvisational sketching exploration.
On the other hand, there are some real-time dynamic AR sketching tools~\cite{suzuki2020realitysketch, saquib2022graphiti, kaimoto2022sketched, suzuki2018reactile}, but they do not allow freehand drawing, limiting the flexibility and generalizability of possible animations.
In contrast, our tool enables both \textit{\textbf{freehand}} and \textit{\textbf{real-time}} AR sketched animation, allowing for more expressive, flexible, and improvisational animation authoring to augment live or recorded real-world interactions through AR.

\begin{figure}[h!]
\centering
\includegraphics[width=\halfwidth]{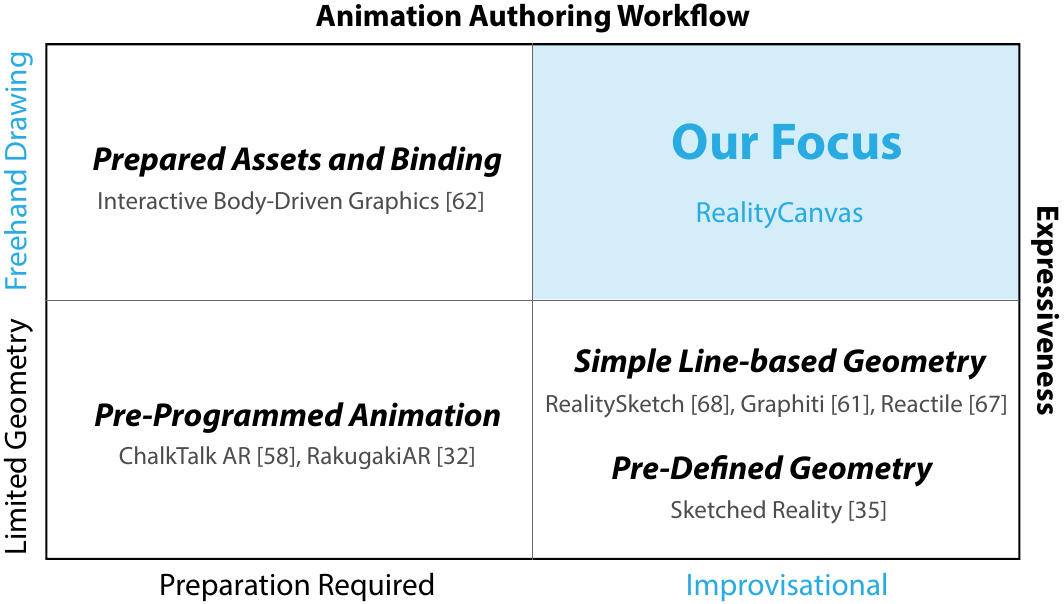}
\caption{Related work of AR sketched animation tools}
~\label{fig:related-work}
\end{figure}



\subsection{Screen-based Sketched Animation Tools}
Apart from AR sketching tools, HCI research also has a long history of screen-based sketching tools (e.g., \textit{SketchPad}~\cite{sutherland1964sketchpad}, \textit{Flatland}~\cite{mynatt1999flatland}, \textit{Silk}~\cite{landay1996silk}, \textit{Teddy}~\cite{igarashi2006teddy}, 
\textit{Electronic Cocktail Napkin}~\cite{gross1996electronic}, 
\textit{ILoveSketch}~\cite{bae2008ilovesketch}).
In particular, recent works have explored real-time interactive sketched animations, which make static sketches dynamic and animated using the power of computation.
For example, \textit{Draco}~\cite{kazi2014draco} explores sketched animation through direct sketching interactions based on kinetic textures.
Such tools have democratized opportunities for animation authoring, which were previously only accessible to professional animators or video editors. 
These tools are used for many applications, such as math and science education (e.g., \textit{MathPad2}~\cite{laviola2004mathpad},  \textit{PhysInk}~\cite{scott2013physink}, \textit{Apparatus}~\cite{schachman2015apparatus}), storytelling (\textit{SketchStory}~\cite{lee2013sketchstory}, \textit{Stop Drawing Dead Fish}~\cite{victor2012stop}), data visualization\removed{s} (e.g., \textit{NpakinVis}~\cite{chao2010napkinvis}, \textit{Data Illustrator}~\cite{liu2018data}, \textit{Transmogrification}~\cite{brosz2013transmogrification}, \textit{DataInk}~\cite{xia2018dataink}), live music performance (e.g., \textit{Megafauna}~\cite{bourgault2021preserving}), and artistic animation authoring (e.g., \textit{Kitty}~\cite{kazi2014kitty}, \textit{Draco}~\cite{kazi2014draco}, \textit{Motion Amplifiers}~\cite{kazi2016motion}).
While these works have expanded possible sketched animation techniques, these interactions are only available on a computer screen. \added{VideoDoodles recently proposed a system to create hand-drawn animations on videos \cite{yu2023videodoodles}. However, their focus is also on video editing and their main contribution lies in a custom tracking algorithm for perspective deformations and occlusions. On the other hand, our focus is mainly on live AR scribble animation. Also, our main contribution lies in the taxonomy and design space, which has not been previously explored.}
Our goal is to bring these \textit{dynamic sketching} techniques into AR sketching tools by leveraging real-world interactions, as opposed to screen-based interactions. 

\subsection{Authoring Tools for Augmented Effects}
Augmented animation effects --- adding visual effects to enhance live or recorded videos --- have been used in many videos on YouTube and TikTok.
Traditionally, creating these effects often requires extensive skills using professional tools such as Adobe Premiere Pro \removed{and}or Final Cut Pro.
For example, tools like \textit{Scribbl}~\cite{scribbl} also provide a means to create scribble animation effects for recorded videos, but these tools only support \textit{frame-by-frame} authoring without any object tracking or binding, which leads to significant time and effort to create animations, limiting real-time and improvisational exploration.
To fill this gap, HCI researchers have explored alternative authoring approaches for such augmented animation effects.
For example, \textit{PoseTween}~\cite{liu2020posetween} lets the user add virtual effects based on human movement to create augmented action videos.
SnapChat's \textit{Lens Studio}~\cite{lens-studio} also allows face augmentation with a simple authoring workflow.
\textit{RealityTalk}~\cite{liao2022realitytalk} also introduces an authoring tool to augment live presentations with embedded graphics.
These tools allow for the creation of impressive augmented animation for various applications, including live storytelling~\cite{saquib2019interactive, liao2022realitytalk}, entertainment ~\cite{liu2020posetween}, education ~\cite{gong2021holoboard}, collaborative discussions~\cite{dillenbourg2013design, kasahara2012second}, data visualization~\cite{chen2019marvist}, and sports training~\cite{chen2018computer, homecourt, sousa2016augmented}.
In these tools, however, the user needs to prepare \textit{assets} in advance to augment the real world. 
In contrast, \system{} lets the user draw these assets through sketching in real-time, which greatly enhances flexibility and improvisational interactions.

%% file: 3-design-space.tex
\begin{figure*}
\centering
\includegraphics[width=\textwidth]{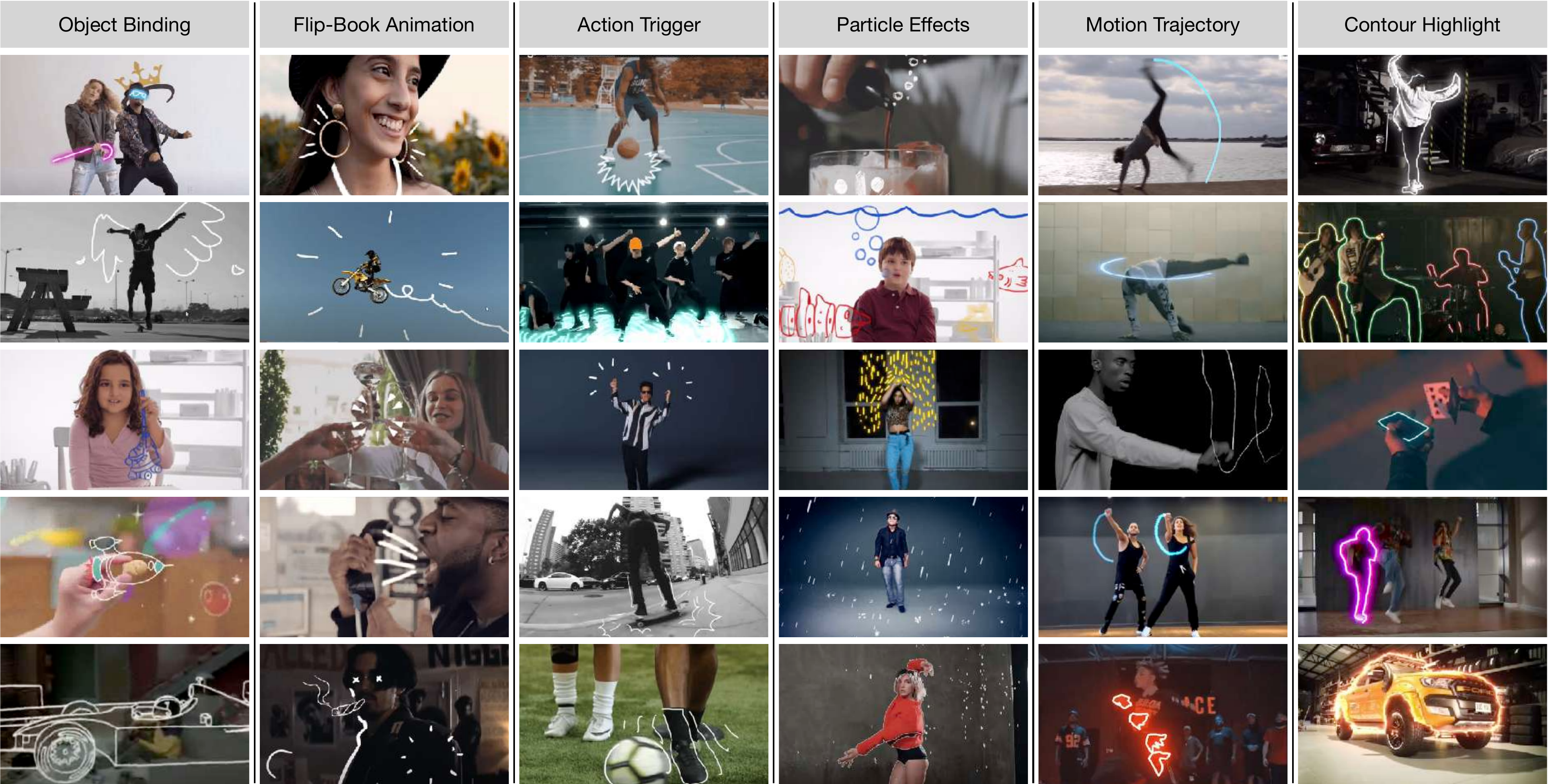}
\caption{Six common scribble animation techniques based on the taxonomy analysis}
\label{fig:design-space}
\end{figure*}

\section{Design Space of Augmented and Responsive Sketched Animation}

\subsection{A Taxonomy Analysis of Scribble Animation Effects}
To better understand the possible design space of augmented and responsive sketched animation, we first collected and analyzed 172 existing video-edited scribble animation examples. 
\textit{Scribble animation effects} are a common video-editing technique to simply add scribble illustrations over a recorded video for each frame.
Because of the flexibility and aesthetics of the scribble effects, such animation techniques are often used in music videos~\cite{bruno-mars, bombay-bicycle-club}, commercial ads~\cite{microsoft, oreo, goldfish-grahams, budweiser}, trailer promotion videos~\cite{dave}, and presentation videos~\cite{ing-direct}.
Currently, these scribble animation effects are created by manual hand-drawing for each frame, but the resulting videos effectively showcase the potential animation techniques for embedded and responsive sketching.

\subsubsection{Corpus and Methodology}
We searched for video examples using popular video and image search platforms (e.g., Google Images, YouTube, Pinterest, Behance, Vimeo, TikTok) by using \textit{``scribble animation''}, \textit{``scribble effects''}, or \textit{``scribble animation effects''} as keywords.
We first collected 172 scribble animation videos, then filtered these videos by focusing only on those with embedded and responsive animations (i.e., remove those where the animation does not interact with or respond to real-world motion), which gave us a total of 120 videos. 
Given the collected videos, one of the authors took screenshots for representative examples for each video and conducted open coding to identify a first approximation of the dimensions by categorizing them. 
After this process, two authors used an online whiteboard (Miro board) to perform systematic coding with individual tagging. Finally, all authors reflected upon the design space and corresponding examples until all agreed on the consistency and comprehensiveness of the categorization.
While we did our best to cover the comprehensive and exhaustive design space, due to the nature of the manual search and taxonomy analysis, we do not argue that our proposed design space is the only way to categorize the existing embedded and responsive sketched animation. 
Rather, our goal is to identify common animation techniques to get insights for ourselves and the HCI community to explore the possibilities of sketched animation for AR. Our results and example videos will be also available on the website~\footnote{\href{ https://ilab.ucalgary.ca/realitycanvas}{ https://ilab.ucalgary.ca/realitycanvas}}.

\subsection{Six Common Animation Techniques}
\subsub{1) Object Binding}
\removed{First, }Object binding is a technique \removed{to bind}for binding sketched elements to a body or physical object, so that the sketches dynamically move according to the corresponding body or object movement.
Examples of object binding include a scribble wing for a human body, scribble glasses or hats for a human face, or scribble \removed{smoke}exhaust for a car. 
The attached sketched objects are mostly static in their shape, so that only the position or orientation changes based on the position of the bound object\removed{ position}.

\subsub{2) Flip-Book Animation}
Flip-book animation is a common animation technique that can be used for many different purposes. In scribble animation effects, many videos use flip-book animation for morphing the shape of the sketches or adding a dynamic effect to existing sketches.
Flip-book animation is often used to combine different animation techniques. 
For example, when combined with object binding, it can create animated object binding effects. 

\subsub{3) Action Trigger}
Action trigger is a one-time animation effect played when a certain action happens. 
For example, when two objects collide, the scribble animation can show\removed{s} a bumping effect to highlight the collision.
Alternatively, such actions can also be bound to different triggered actions.
Possible actions we have observed are hand-clapping, kicking, touching down, hand-waving, and stomping, among many others.

\subsub{4) Particle Effects}
Particle effects are a technique to create repeated animation with many spawned elements.
For example, some videos use scribble particle effects to effectively create raindrops, wind flow, or smoke animations.
In contrast to action-trigger, particle effects are \removed{supposed}intended to be repeated effects.

\subsub{5) Motion Trajectory}
Motion trajectories are used to highlight a certain movement of the body or object.
By showing the path trajectory, the video can make physical motion more visible and expressive.
Such motion trajectory is often used to make human actions stand out, \removed{like with}as in dancing and sports videos.
Most of the motion trajectories use a simple path animation that shows the afterimage of the body.
Some other examples create more expressive motion effects by using a morphing animation. 

\begin{figure*}
\centering
\includegraphics[width=\textwidth]{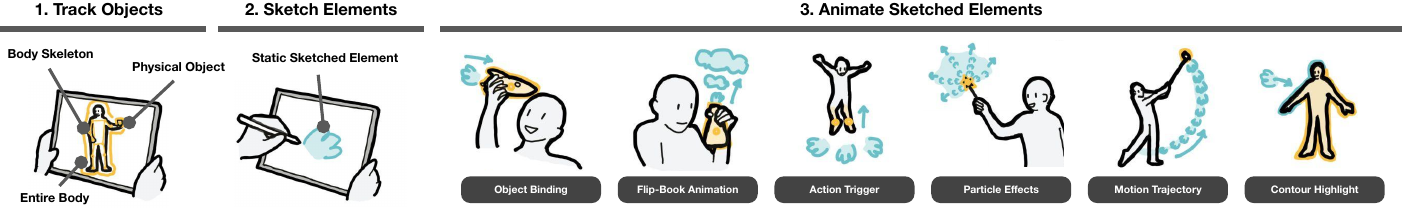}
\caption{Authoring Workflow of \system{}}
\label{fig:overview}
\end{figure*}

\subsub{6) Contour Highlight}
Finally, many scribble animations use an animated contour line to trace a body or object.
This allows a \removed{certain}selected object or body part to stand out by highlighting its contour.
When the body posture changes or the object moves, these contour lines keep following to fit the bound object. 
Most of the contour highlight lines are static, in which the entire contour is covered by a single line, but some examples leverage an animated line, in which a partial line moves across the contour of the object. 

%% file: 4-implementation.tex
\section{RealityCanvas}
\subsection{Overview}
Based on the above design space, we designed and developed \system{}, an AR sketching tool to augment a captured real-world interaction through responsive scribble animation effects. 
As a basic setup, \system{} uses a mobile device and a standard RGB camera. Therefore, the user can sketch elements with a finger, a pen, or a mouse on smartphones, tablets, and laptop computers.
The \system{} system uses computer vision for object tracking and interaction detection, allowing it to work with either live or recorded videos. 
For recorded videos, users can pause the video to sketch and define animations, which then start animating when the video is resumed. For live videos, users can embed animations onto a real-time video stream. In our setup, a performer can see the live sketched animation through an external monitor connected to the tablet, similar to~\cite{saquib2019interactive, liao2022realitytalk}, and collaborate with the user to create a live performance that can later be saved as a video. All sketched elements are displayed as 2D objects on a screen, without any depth effect based on camera movement. 
The system is developed using web technology and a live demo is available on our website~\footnote{\href{https://ilab.ucalgary.ca/realitycanvas}{https://ilab.ucalgary.ca/realitycanvas}}. To make AR sketched animation accessible to everyone, we will release the source code as an open-source project~\footnote{\href{https://github.com/ucalgary-ilab/realitycanvas}{https://github.com/ucalgary-ilab/realitycanvas}}.

\subsection{Two Key Design Goals}
\system{} is designed with two key goals in mind: \textit{expressiveness} and \textit{improvisation}. \textbf{\textit{Expressiveness}} refers to the range of animation effects that the system can provide, and we have designed six types of sketch animation effects (A1-A6 below) to cover a wide range of expressions, informed by the design space analysis of existing scribble animations. This allows for open-ended animation drawing, which was not possible with prior systems~\cite{suzuki2020realitysketch}. 
\textbf{\textit{Improvisation}}, on the other hand, enables users to spontaneously create animation \textit{in real-time} without prior preparation or planning, as opposed to prior systems~\cite{saquib2019interactive, perlin2018chalktalkvrar}. This is achieved through both freehand drawing and immediate response, which are crucial not only for faster creation but also for unexpected discoveries through playful exploration and experimentation. This design is guided by Victor's principle (\textit{``creators need an immediate connection to what they are creating''}~\cite{victor2012inventing}), and we believe that the support of improvisation enables the users to engage with this new medium to generate new and original ideas they may not have thought of before.

\subsection{Authoring Workflow}
At a high-level, \system{} provides the following three-step workflow: 1) object selection and tracking, 2) sketch elements, and 3) animate sketched elements.

\ \\
\subsub{Step 1. Object Selection and Tracking}
To create embedded and responsive animations, the sketched elements need to be tightly coupled with the body, objects, and/or environments. 
Thus, capturing and tracking an object is the first step to making the sketches responsive to physical motion.
To track an object, the user can simply pause the video input or ask the performer to hold still. Then, the user can select a point of an object (based on color tracking) or a part of a body (based on skeleton tracking).
Once selected, the system \removed{keeps tracking} overlays a green dot  on the selected point to indicate the tracked point on the body or object.

\ \\
\subsub{Step 2. Sketch Elements}
Once the user selects a tracking point, they can start drawing a sketched element.
The user can simply sketch any shape by freehand drawing in the scene.
The user can also change the color, thickness, and opacity of the \removed{pen} stroke when drawing.

\ \\
\subsub{Step 3. Animate Sketched Elements}
The key feature of \system{} is the animation process. 
Informed by the design space exploration, the system provides six different ways to animate the sketched elements through direct manipulation.
\begin{enumerate}
\item[\textbf{A1.}] \textbf{Object Binding:} Bind the sketched element to a \removed{tracking point and move its position} tracked point on an object or body part
\item[\textbf{A2.}] \textbf{Flip-book Animation:} \removed{Add new} Add multiple frames of sketches to create a flip-book animation effect
\item[\textbf{A3.}] \textbf{Action Trigger:} Define the trigger to specify when the sketched elements appear
\item[\textbf{A4.}] \textbf{Particle Effects:} Draw a line to spawn many cloned elements
\item[\textbf{A5.}] \textbf{Motion Trajectory:} Specify a tracked point to show the path by cloning elements
\item[\textbf{A6.}] \textbf{Contour Highlight:} Select the object to highlight its contour with a sketched line
\end{enumerate}
In the following, we describe each sketched animation technique in more detail.

\subsection*{A1. Object Binding}
\begin{figure}[h!]
\centering
\includegraphics[width=\halfwidth]{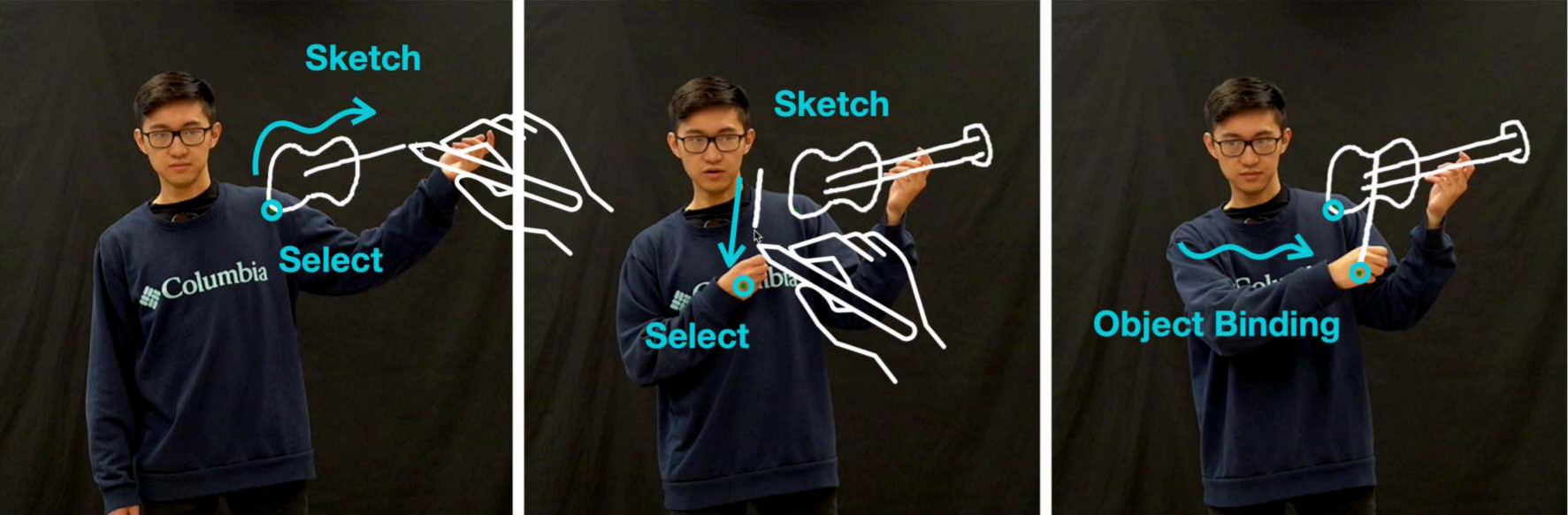}
\caption{Object Binding}
\label{fig:object-binding}
\end{figure}

Once the user selects an object to be tracked, the user can start sketching an element.
By default, the sketched elements are automatically bound to the selected tracking point, so that the sketched element starts moving when the tracked object moves. 
For example, in Figure~\ref{fig:object-binding}, the user selects a shoulder as a tracking point, then draws a violin around the tracked point, so that the sketched violin starts moving when the shoulder moves.
In the same way, the user tracks a right hand and binds a sketched violin bow; then, the bow also moves based on the user's hand movement.

This animation technique is useful for \textit{\textbf{object binding}}, in which the sketch is bound to a body or object to augment the real-world scene. 
In this simple object binding, the sketched element itself is static and only its position moves and its orientation \removed{or}and scale do\removed{es} not change.

\subsection*{A2. Flip-Book Animation}
\begin{figure}[h!]
\centering
\includegraphics[width=\halfwidth]{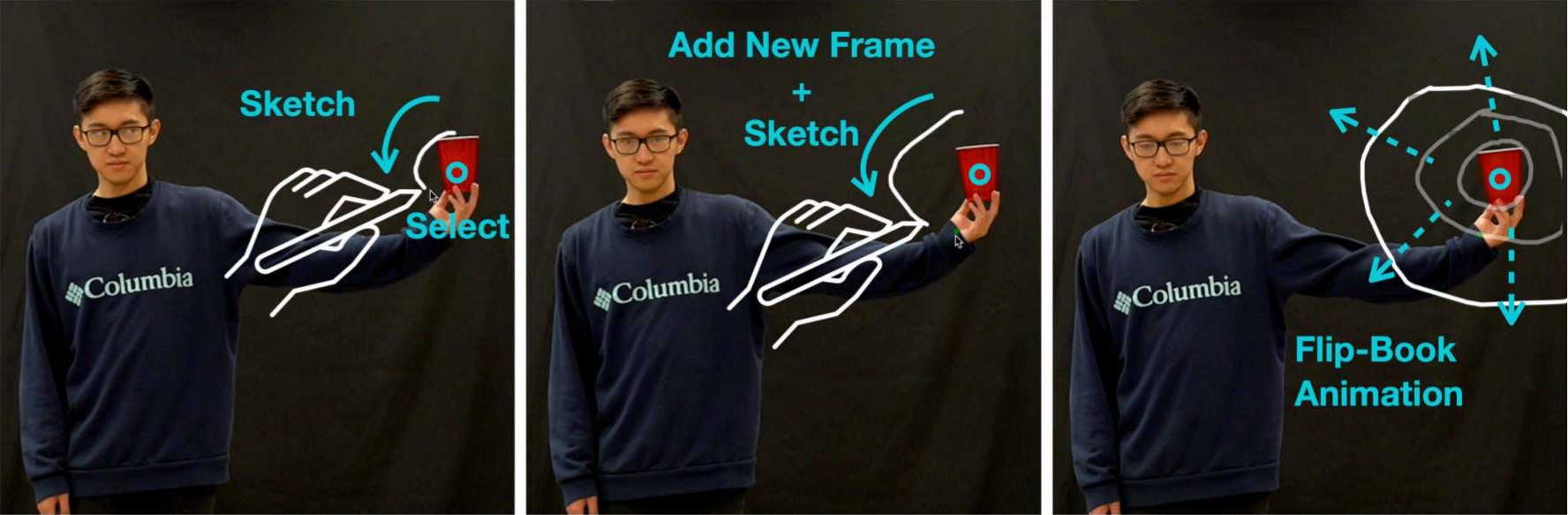}
\caption{Flip-Book Animation}
\label{fig:flip-book-animation}
\end{figure}

To make more expressive animation, the system allows the user to create \textit{\textbf{flip-book animation}} effects.
In this animation technique, the user can draw an additional sketch as a new frame, and the system shows these sketches one by one to create an animation. 
For example, in Figure~\ref{fig:flip-book-animation}, the user first selects a red cup, then draws a circle around the selected point
while the performer holds still in live mode. Similarly, the user can pause the recorded video at a specific frame to sketch the circle. 
Then, the user selects the add a new frame button so that they can draw the next shapes for frame-by-frame animation. 
Once the user finishes drawing all frames, the user clicks the save button to apply the flip-book animation. 
In this way, the user can create an animated circle whose size appears to change.
The drawn animation is also bound to the tracked object, so that the animated circle moves when the tracked object (red cup) moves.

\subsection*{A3. Action Trigger}
\begin{figure}[h!]
\centering
\includegraphics[width=\halfwidth]{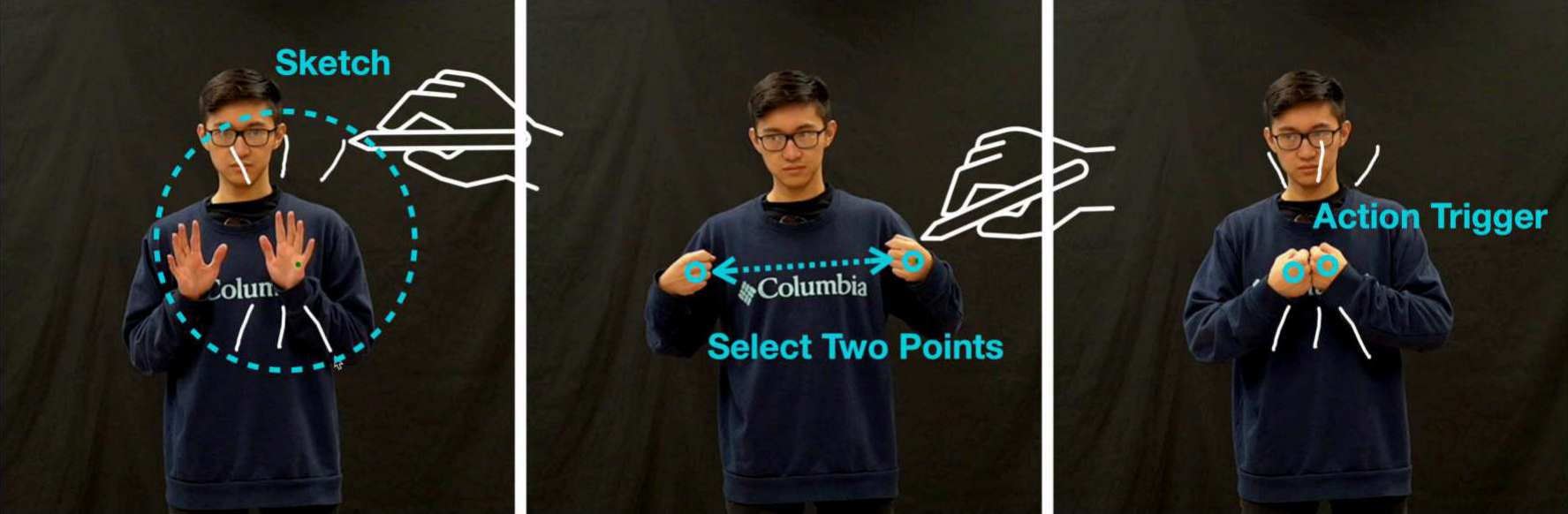}
\caption{Action Trigger}
\label{fig:action-trigger}
\end{figure}

The above two animation techniques are always-visible animations.
On the other hand, \textit{\textbf{action trigger}} animations allow the user to create animations that appear only when a certain action occurs. 
To use this feature, the user can simply track multiple points and specify the user-defined trigger action.
In our system, the user can define the trigger based on the distance between two points.
For example, in Figure~\ref{fig:action-trigger}, the user first tracks left and right hands, and then the user draws a bumping sketched effect.
Once the user clicks the trigger button, then the system continuously tracks the distance between the two points. 
By default, the system triggers the sketched bumping effect when the distance between the two points decreases below a threshold of a user-defined pixel distance \removed{(based on a pixel distance on a screen)}.
The user can adjust the trigger parameters using a slider to determine the distance threshold value and a button to change the triggering direction. By toggling the direction, the user can trigger the action when the distance between two points increases rather than decreases. This allows for greater flexibility in defining the trigger conditions.


By combining this with the above flip-book animation effect, the user can create even more expressive animations based on human action.
For example, Figure~\ref{fig:teaser} illustrates how the user can show a water splashing effect when the actor stomps the ground.
In this case, the user can select the left foot and the ground as triggering points using both body tracking and object tracking, then draw flip-book animations \removed{multiple frames} as trigger action effects. 
In this way, the user can create a variety of animations based on the user's action.

\subsection*{A4. Particle Effects}
\begin{figure}[h!]
\centering
\includegraphics[width=\halfwidth]{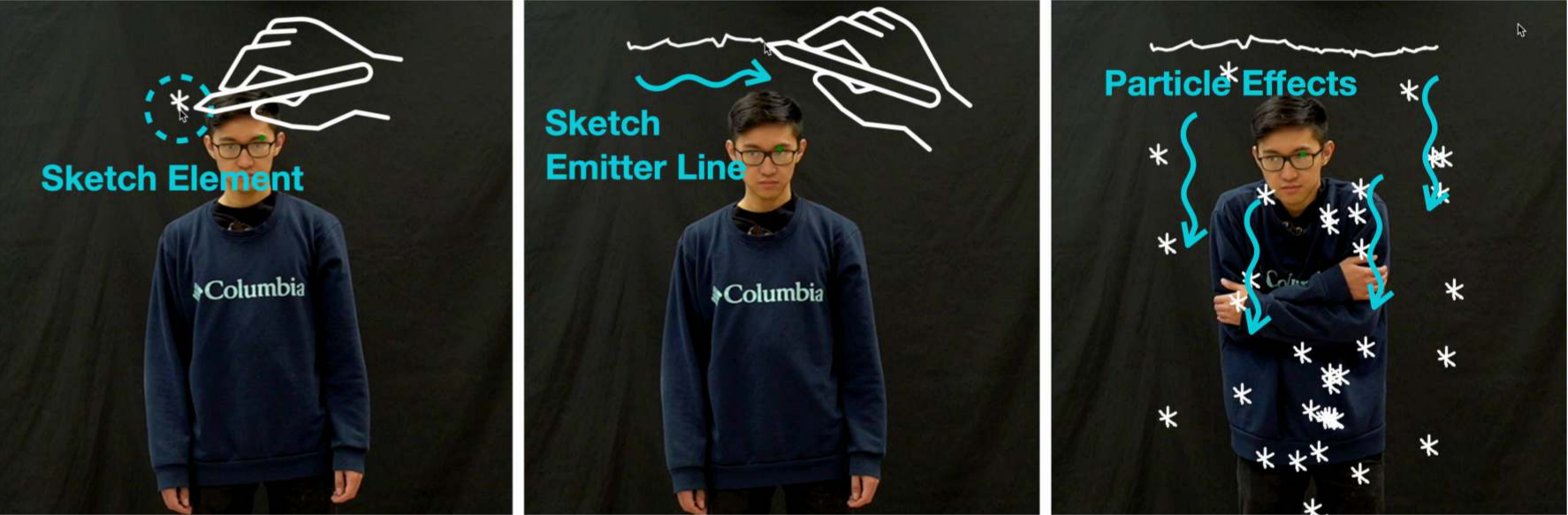}
\caption{Particle Effects}
\label{fig:particle-effects}
\end{figure}

The system also lets the user create \textit{\textbf{particle effects}} through improvisational sketching interactions.
To create particle effects, the user first sketches an element of the particle, such as a raindrop in Figure~\ref{fig:teaser} or snowflake in Figure~\ref{fig:particle-effects}.
Then, the user selects the tracked object and clicks the emit button, which allows the user to draw an \textit{emitting line} to spawn new elements for particle effects.
As shown in Figure~\ref{fig:particle-effects}, the same snowflake sketches are randomly emitted from the emitter line for the snow effects. The system offers two options for customizing the particle effect parameters: 1) the user can draw a \textit{motion path} to specify the path each particle follows, and 2) the user can change the \textit{speed} of particle movement using a slider in a menu. These options were inspired by the interactions presented in \textit{Draco}~\cite{kazi2014draco}. 

Our interaction technique is inspired by Draco~\cite{kazi2014draco}'s kinetic texture and our tool also supports similar interactions, such as allowing the user to draw a trajectory of the emission so that the user can control the path the emitted particles follow.
But unlike Draco, which focuses on screen-based interaction, the drawn emitter line in \system{} is bound to the selected physical object so that if the object moves, the emitter line follows along. This enables a broader interaction space, such as a magic wand particle effect from a physical stick or showing an air flow bound to a dryer.

\subsection*{A5. Motion Trajectory}
\begin{figure}[h!]
\centering
\includegraphics[width=\halfwidth]{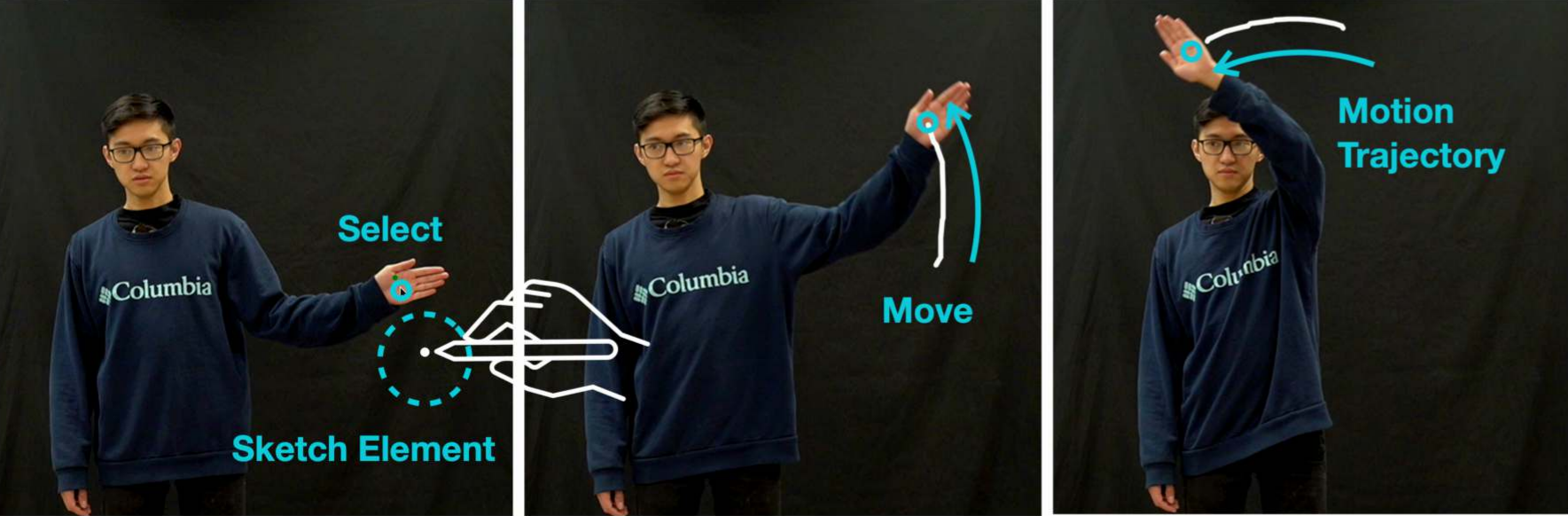}
\caption{Motion Trajectory}
\label{fig:motion-trajectory}
\end{figure}

In object binding, the sketched element simply moves and follows the tracked point, but by duplicating the sketched element for each tracked position, the system can also create a \textit{\textbf{motion trajectory}} like ghost effect of the sketched elements.
For example, in Figure~\ref{fig:motion-trajectory}, the user selects a left hand as a tracked point and sketches a simple dot. 
When the user clicks the motion button, the dots become a line to show the trajectory of the hand movement.
With a certain time period (by default 30 elements), the cloned elements start disappearing, so that the motion trajectory disappears with a certain length.

This animation technique is useful when the user wants to visualize the trajectory of a tracked object, such as with a dancing motion or throwing a basketball.
The user can also change the parameters of the motion trajectory, such as how fast the elements should disappear, or the opacity or scale of each cloned element.
By leveraging these parameters, the user can also create more expressive motion effects.

\subsection*{A6. Contour Highlight}
\begin{figure}[h!]
\centering
\includegraphics[width=\halfwidth]{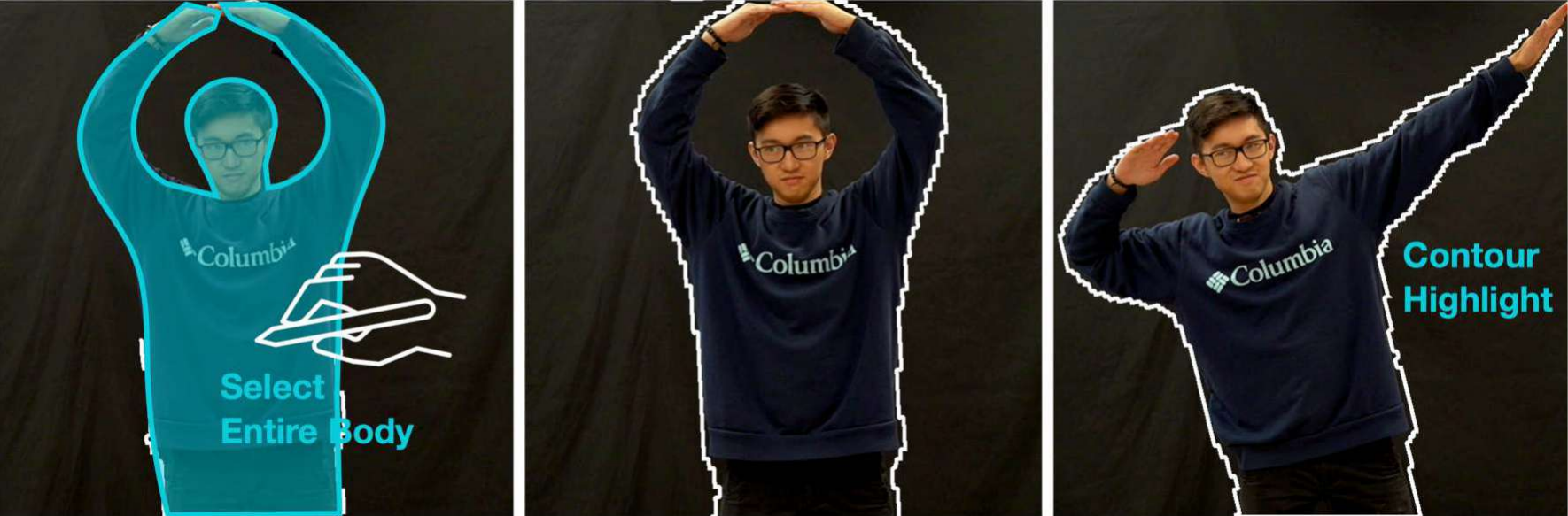}
\caption{Contour Highlight}
\label{fig:contour-highlight}
\end{figure}

Finally, the system lets the user create \textit{\textbf{contour highlight}} effects for a tracked body or object.
To do so, the user first selects the contour button and then taps the body or object the user wants to highlight.
By default, the enclosed object or body is highlighted with a drawn line.
This line can be changed according to the body motion or object movement, as \system{} always tracks and displays the outermost object contour seen in Figure~\ref{fig:contour-highlight}.
In a similar way, the user can also create animated contour highlights by selecting the appropriate option. 
For example, in Figure~\ref{fig:contour-highlight}, the user can also show an animated contour line around the body. 

The user can also fill in the enclosed object. 
For example, the user can change the fill color of the body by combining the action-trigger effect. 
The user first selects the foot and defines the trigger based on the foot location, the user can create the highlighting effect along with the animated elements.

\subsection{Implementation Details}

\subsubsection{Sketching Interface}
\system{} uses HTML Canvas as the main sketching interface.
Since most of the sketched elements are 2D, we decided to embed sketches onto the 2D canvas screen. 
All sketched elements are drawn in SVG format on the canvas. 
The system also uses Konva.js and Anime.js as supporting graphics and animation libraries, respectively.
All elements are calculated and rendered as 2D objects.
For example, the distance between two tracked points is based on the pixel value on the screen, rather than the physical distance. 
Therefore, \removed{the }depth effects and \removed{or} 3-dimensional rotation \removed{is}are currently not supported (although the user can still draw sketches which look 3-dimensional using perspective effects).

\begin{figure}[h!]
\centering
\includegraphics[width=0.6\halfwidth]{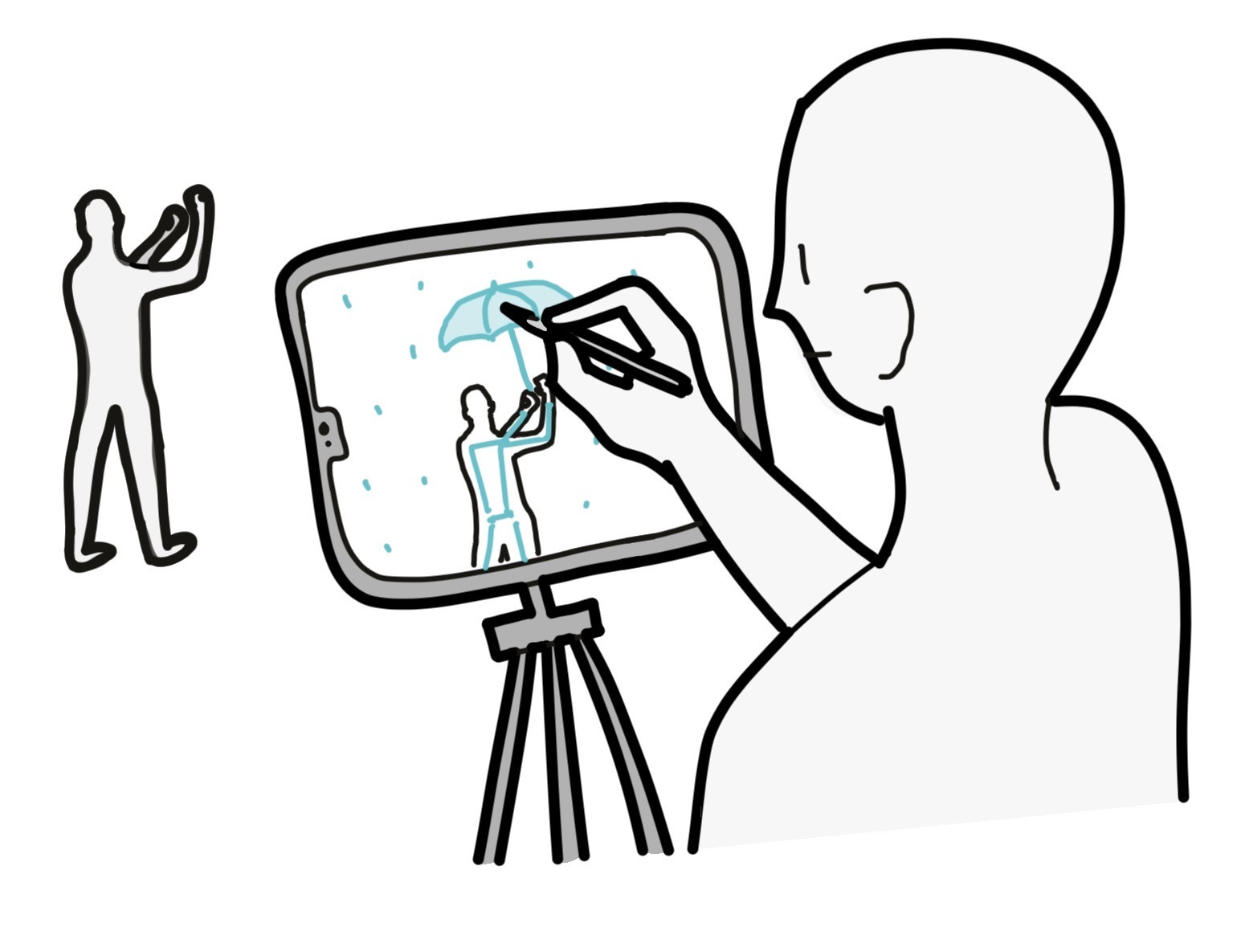}
\caption{\added{In the live mode, the tablet is setup on a tripod at a similar angle to a canvas on an easel which helps the tablet capture the real world and allows the user to sketch on the tablet.}}
\label{fig:setup}
\end{figure}

\subsubsection{Object Tracking}

For object tracking, we use a simple color-based tracking method that has been shown to provide fast and reliable real-time tracking for AR animation applications~\cite{suzuki2020realitysketch}.
In our current implementation, the system uses OpenCV JS to track objects based on their RGB values, with a certain threshold to filter out pixels outside of the specified range.
The system first obtains the RGB value based on the selected X-Y point by sampling the pixel at the corresponding coordinates in the video frame.
The system then filters out all pixels outside the specified RGB range. Then the system identifies the largest continuous contour as the tracked object. \removed{captures the largest contour object given the}. The RGB value range is ($r\pm10$, $g\pm10$, $b\pm10$), where ($r, g, b$) is the sampled RGB value of the selected point.
Given the largest contour, we obtain the center point of the object, which is used to determine the object's location in the video frame. 
This information is then used to track the object's movements over time and update its position on the screen accordingly. We evaluated the robustness of our color tracking system \added{by tracking a reflective object (a fake apple) and a non-reflective object (a tennis ball) under four different lighting conditions:}

\added{
    \begin{enumerate}
        \item Dim artificial lighting: Simulated using a single light source in the living room.
        \item Bright artificial lighting: Achieved by activating four light sources in the living room.
        \item Dim natural lighting: Simulated by creating subdued sunlight conditions in the living room.
        \item Bright natural lighting: Achieved by exposing the living room to direct sunshine.
    \end{enumerate}
}
\added{ The success of the tracking was determined by the accurate movement of a green dot, indicating successful object tracking, while failure was identified when the green dot failed to follow the object.} We found that the system performs more accurately in bright lights than in dim lighting. In dim artificial light, the system successfully tracked the object 5 times and failed to track it 5 times. However, in bright artificial light, it successfully tracked the object 8 times and failed to track it only 2 times. We also discovered that the mesh of an object can impact the accuracy of color tracking in bright light. Objects with reflective mesh in bright lighting often result in a high false negative rate. Our system performed best when the objects had distinct solid colors.

\subsubsection{Body Tracking}
In our current implementation, the system tracks body parts based on MediaPipe.
MediaPipe provides 33 body points based on the human skeleton. When the user clicks a point on the screen, the system identifies and \removed{keeps tracking}continuously tracks the closest body point among the 33 body points based on X-Y value.  
\removed{Additionally, we also use the same MediaPipe library for hand skeleton and facial anchors to track points on the hand and face when visible.} 
For body segmentation, we used MediaPipe and OpenCV JS to obtain the masked shape of the body. 
Based on the masked image, we convert the body shape into the simplified SVG path for contour line animation in each frame when the contour highlight is enabled.

\subsubsection{Prototype Setup}
Since all of the components including the color tracking and body tracking are client-side systems, we do not have any server to run the system.
During the prototyping phase, we tested with several devices, including Pixel 3 XL, Pixel 6, Linux Desktop, Windows machine, and Android tablet. (Due to the Safari browser's restriction, the iPhone/iPad did not work, although this problem should be resolved in the future.)
For the demo and user evaluation, we used Samsung Tab S8 Ultra Android Tablet (Display: 14.6-inch, CPU: Qualcomm SM8450 Snapdragon 8 Gen 1, GPU: Adreno 730, RAM: 12GB) with a built-in stylus pen.
In this setup, the system achieved the targeted 60 FPS at all times \added{with the tablet stock camera} during our testing for both color tracking and body tracking. \added{However, we observed a decrease in frame rate when using an external high-resolution camera.}

%% file: 5-application.tex
\begin{figure*}
\centering
\includegraphics[width=\textwidth]{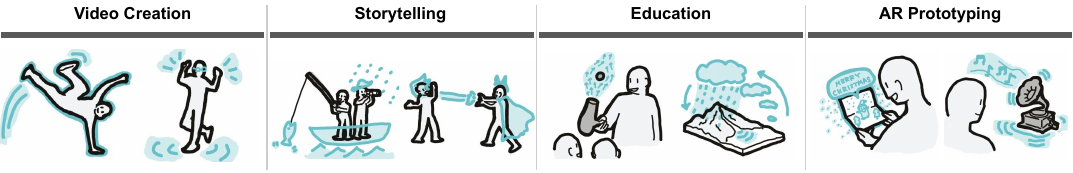}
\caption{Applications Scenarios}
\label{fig:applications}
\end{figure*}

\section{Applications}

\subsection{Social Media Video Creation}
First, our system allows users to quickly create videos for social media platforms such as TikTok, YouTube, and Instagram. Currently, video creation and editing with these animation effects \removed{have a tedious workflow} require specific skills and non-trivial amounts of time. People can use our system to record their videos with animated effects much more easily. For example, the user can shoot a dancing or music video with real-time scribble animation effects, such as \textit{motion trajectory} of dancing movement, \textit{contour highlight} to make a person stand out, and \textit{action trigger} to highlight the claps or steps. 

\subsection{Augmented Storytelling}
Using our system, the user can also easily create augmented \added{stories}~\cite{saquib2019interactive, liao2022realitytalk}.
For example, by using animated \textit{object binding}, the user can augment the body with sketched illustrations and animation, like adding a Superman costume or pirate features. In this scenario, the user can sketch on another user’s performance, which can be streamed to a large screen in real time so that the performer can see the current performance. 
Such augmented storytelling would greatly expand, for example, children’s storytelling experience with dynamic effects.


\subsection{Classroom Education}
The user can also add dynamic visual annotations to a physical phenomenon to provide better explanations for science or physics classrooms. In this scenario, similar to the storytelling scenario, the teacher can stream the screen view to a large display to make the animation visible to the entire classroom. For example, by using \textit{particle effects}, the user can visualize the airflow \added{around} a levitated ping pong ball. The user can explain how water circulates on Earth by using an animated sketch of rain and clouds with \textit{particle effects} by augmenting a physical model such as terrains. Similar to \textit{RealitySketch}~\cite{suzuki2020realitysketch} and \textit{HoloBoard}~\cite{gong2021holoboard}, such augmentations can enhance the learning experience for classrooms. 

\subsection{Prototyping AR Applications}
Finally, the \system{} system can also be used to create an interactive and animated prototype for AR applications. Similar to existing AR prototyping tools~\cite{leiva2021rapido, leiva2020pronto, nebeling2019360proto, nebeling2018protoar}, embedded sketches allow the user to simulate AR experiences for low-fi prototypes quickly. For example, the user can quickly create an animated AR Christmas card using \textit{object binding}. With \textit{action trigger}, the user can create some interactive AR linked to an action, like the opening of the card. The user can also sketch the animation for augmented music visualization like AR Music Visualizer~\cite{tanprasert2022ar} to test how different effects can enhance the music experience. 

%% file: 6-user-study.tex

\section{Usability Study}
\subsection{Method}
We evaluate our system with two user studies: a usability study and expert reviews. The first study's goal is to evaluate our system's usability.
To do so, we recruited twenty participants (14 male, 6 female, age: 19 - 29) from our local community.
The usability study consists of the following two tasks.

\subsubsection{Standardized Task} 
We first demonstrated all of the basic functionalities. We then showed an animation video (Figure~\ref{fig:teaser}) we created with our system. The video featured all the functionalities. The animation video had the following elements: 1) \textit{object binding} between hands and umbrella, 2) \textit{flip-book animation} for the triggered water splash, 3) \textit{action trigger} based on the stomp, 4) \textit{particle effects} with rain drops, 5) \textit{motion trajectory} of the hand movement, 6) \textit{contour highlight} of the whole body. 
We asked the participants to recreate this animation without instructing them how to make it.
The goal of this task is to let the user use all of the functionalities to measure each feature's usefulness and understandability.

\subsubsection{Exploration Task}
Once the user finished the identical task, we moved on to the exploration task. We asked the participants to create their own sketched animation from their imagination. 
For inspiration, participants were shown existing examples of scribble animations samples from the corpus we collected for our taxonomy analysis as well as many animation videos created by the authors. Each participant was asked to think of a scenario that involves one or more objects driven by a human action or object movement. Then one of the authors performed the actions for the participant while the participant added sketches on top of it by themselves. The goal of this task was to measure the expressiveness and flexibility of our system through open-ended animation scenarios.

\ \\
We designed these tasks as \textit{``usage evaluation''} based on the HCI toolkit evaluation strategy~\cite{ledo2018evaluation}.
Instead of a controlled experiment, we chose to conduct the usage evaluation because there was no clear baseline to compare with \system{}.
For example, existing video-editing tools have a completely different workflow, so we cannot compare them with our system. 
Therefore, for the usability study, we focused on the usage evaluation for the end user. 
On the other hand, we also conducted an expert review to compare our approach with existing video-editing or animation techniques from an expert's point of view. 

Both tasks were conducted in a research lab.
All of the participant's performances were recorded for objective measurements (e.g., task completion time).
After the session, we asked the participants to provide feedback through an online questionnaire. 
In total, the study took approximately 60 minutes, and participants were compensated \$10 CAD.

\begin{figure}[h!]
\centering
\includegraphics[width=0.45\textwidth]{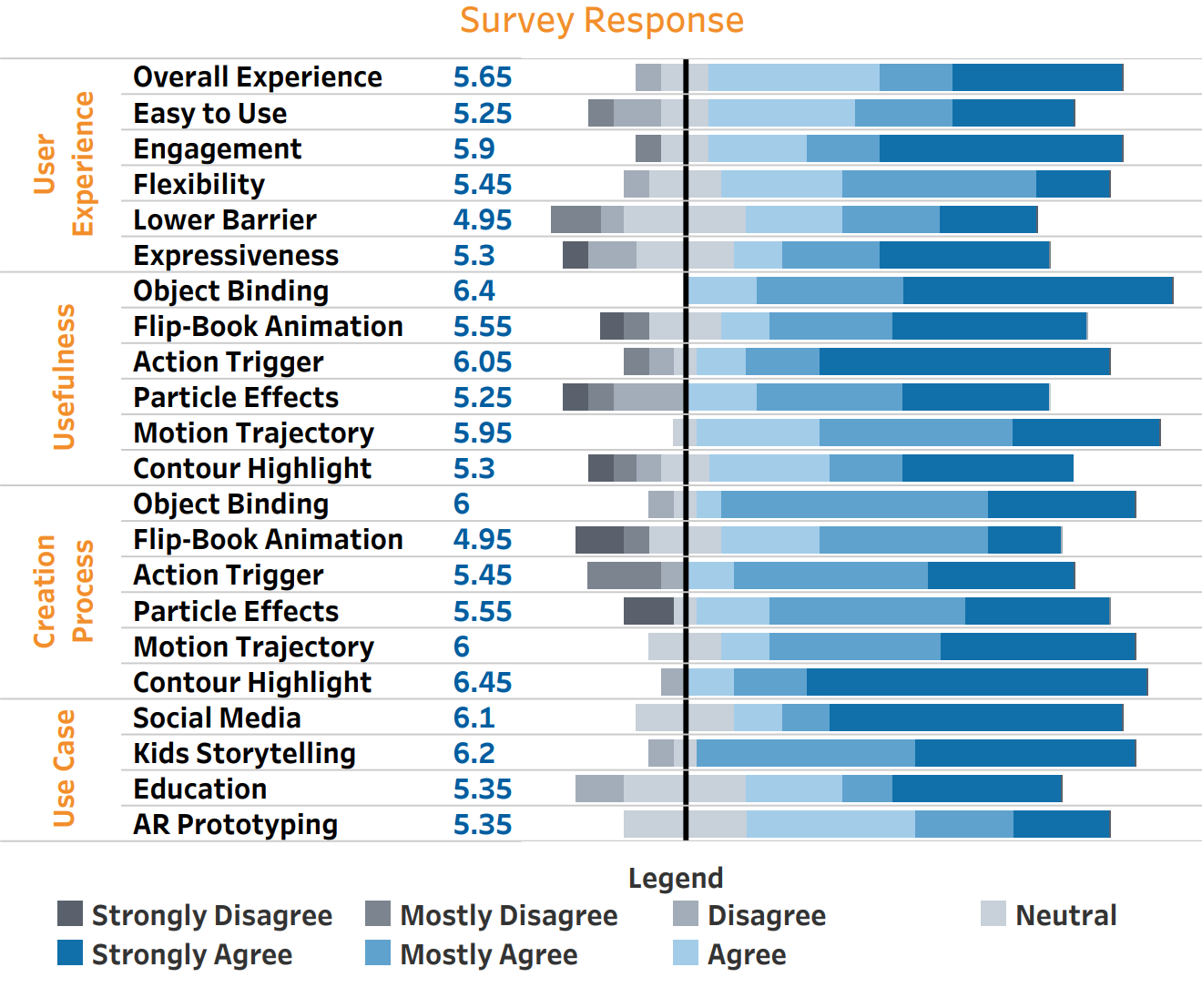}
\caption{Usability Study Results - A graph summarizing the 7-point Likert scale responses for 20 participants.}
\label{fig:usability-result}
\end{figure}

\begin{figure*}[h!]
\centering

 \begin{subfigure}[t]{.15\linewidth}
    \centering
    \includegraphics[width=\linewidth]{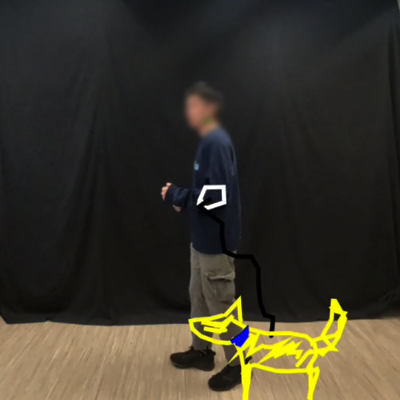}
    \caption{Walking a dog}
  \end{subfigure}
  \begin{subfigure}[t]{.15\linewidth}
    \centering
    \includegraphics[width=\linewidth]{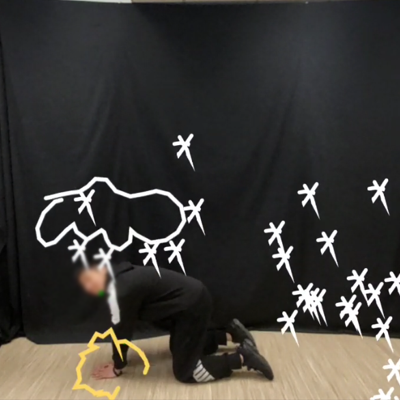}
    \caption{Cloud and snowfall}
  \end{subfigure}
  \begin{subfigure}[t]{.15\linewidth}
    \centering
    \includegraphics[width=\linewidth]{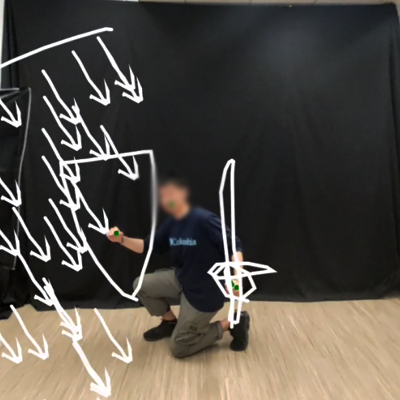}
    \caption{Arrows and a shield}
  \end{subfigure}
   \begin{subfigure}[t]{.15\linewidth}
    \centering
    \includegraphics[width=\linewidth]{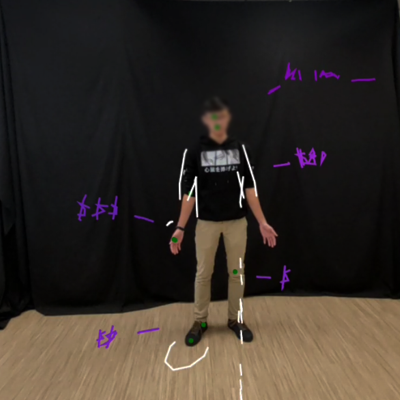}
    \caption{Body annotations}
  \end{subfigure}
  \begin{subfigure}[t]{.15\linewidth}
    \centering
    \includegraphics[width=\linewidth]{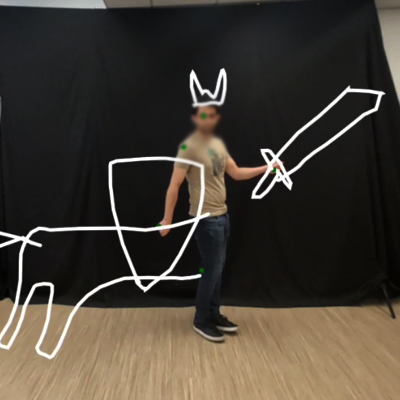}
    \caption{Pretending to be a centaur}
  \end{subfigure}
  \begin{subfigure}[t]{.15\linewidth}
    \centering
    \includegraphics[width=\linewidth]{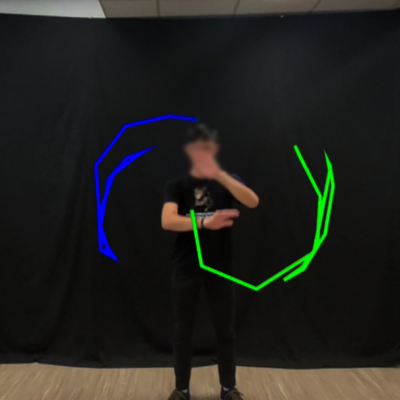}
    \caption{Playing Tai Chi}
  \end{subfigure}
  \begin{subfigure}[t]{.15\linewidth}
    \centering
    \includegraphics[width=\linewidth]{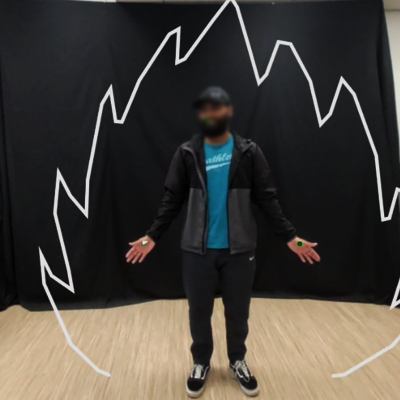}
    \caption{Bursting flame}
  \end{subfigure}
  \begin{subfigure}[t]{.15\linewidth}
    \centering
    \includegraphics[width=\linewidth]{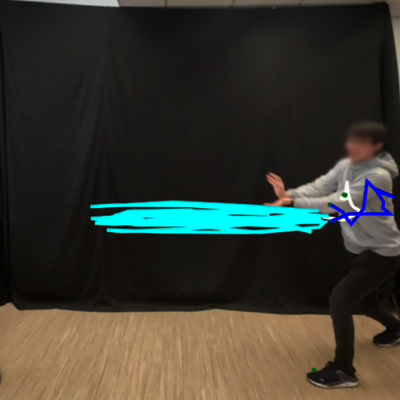}
    \caption{Shooting energy blast}
  \end{subfigure}
  \begin{subfigure}[t]{.15\linewidth}
    \centering
    \includegraphics[width=\linewidth]{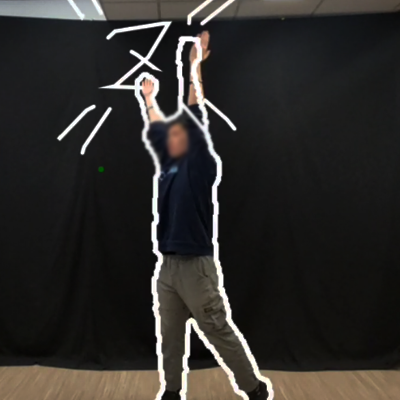}
    \caption{Flashing text}
  \end{subfigure}
  \begin{subfigure}[t]{.15\linewidth}
    \centering
    \includegraphics[width=\linewidth]{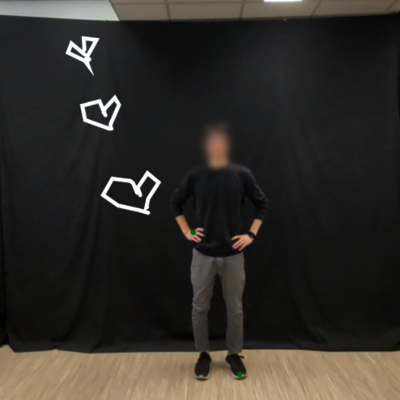}
    \caption{Floating hearts}
  \end{subfigure}
  \begin{subfigure}[t]{.15\linewidth}
    \centering
    \includegraphics[width=\linewidth]{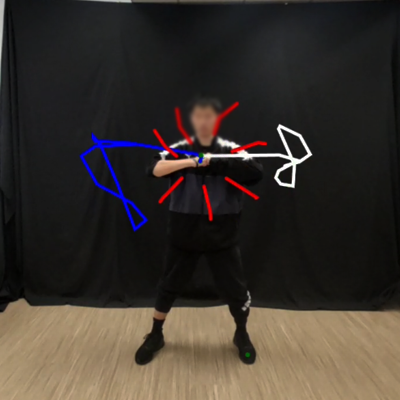}
    \caption{Fist bumping}
  \end{subfigure}
    \begin{subfigure}[t]{.15\linewidth}
    \centering
    \includegraphics[width=\linewidth]{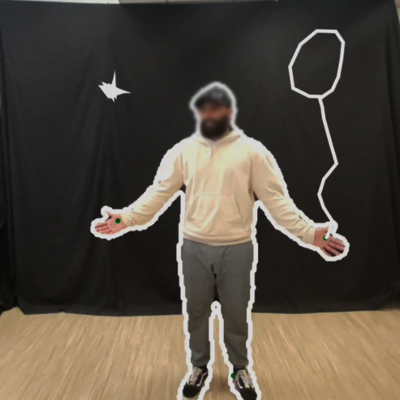}
    \caption{Holding a balloon}
  \end{subfigure}
\caption{Exploration study scenes created by participants}
\label{fig:exploration-study}
\end{figure*}

\subsection{Results}
\subsubsection*{\textbf{1) Overall Experiences}}
Figure~\ref{fig:usability-result} summarizes the 7-point Likert questionnaire response of the usability study.
Overall, participants responded positively to the user experience (5.65/7). 
\textit{``P7: The entire system was quite neat and fun to use. When watching the examples, I imagined myself re-creating them.''}
The participants also responded positively about the interface and workflow.
From the questionnaire results, the participants found the system easy to understand (5.25/7), and from the comments, many participants mentioned that the interface was self-explanatory (P2, P5) and easy to learn (P8, P12).
\textit{``P5: The interface and the workflow were simple enough to learn that little explanation was required.'' }

Most participants did not have animation or video creation experience, but the participants felt that the system could be easy to use even for novice users. 
The participants generally agreed that the system would lower the barrier to creating AR animation (4.95/7).
\textit{``P15: It was very easy to pick up the tools and nuance of the system. Further, I can imagine many ways in which both amateurs and professionals can make interesting things using it.'' } 
For the standardized task, all participants could complete the given task without any explicit assistance from the researcher.
For the task completion time, the participants finished the standardized task on average in 3.25 min (Min: 1.48 min, Max: 8.2 min, SD: 1.58 min). 

While the participants did not see any issues with the workflow, some complained about the system's performance.
For example, the participants commented that the animation was not smooth (P14) and sometimes lagged (P9, P11).
Also, the participants commented that the tracking accuracy could be improved (P18). 
Additionally, the participants wished for finer-grained body and object tracking (P12, P19). 

\subsubsection*{\textbf{2) Quality and Expressiveness}}
One of the goals of our system is to increase the expressiveness and flexibility of sketched animations with freehand drawing.
From the questionnaire results, participants found the system flexible (5.45/7) and agreed that the quality and expressiveness of the created result are high (5.3/7).
For example, many participants found the system flexible (P2, P10, P18, P20) to create various animations,
\textit{``P20: Based off of what I tried to do, it has enough features to support different creative ideas.''} 
The participants also positively commented on the expressiveness of freehand drawing. 
\textit{``P7: As mentioned previously, I think once one has a better understanding of the system (more than ~20 mins), there are endless possibilities.''}
For the exploration task, the participants created various animations with 5.24 min on average (Min: 2.21 min, Max: 18.52 min, SD: 4.87 min) \added{and median 8.93 min}. 
Figure~\ref{fig:exploration-study} illustrates the example animations created by the participants. 
Among the 20 scenarios, we observed some common patterns such as dancing highlights, using superpowers, and equipping weapons.

\subsubsection*{\textbf{3) Creation Process and Improvisation}}
During the creation process for exploration tasks, we observed that most participants did not have a clear idea of what they wanted to create at first. Instead, they developed their narratives by drawing a simple shape and experimenting with supported animation features. For instance, in the case of the arrows and shield example (c), the participant first drew a simple arrow symbol and played with the emitting line. As they did so, the arrow began to look more like arrows from an enemy, which led to the story of a fighting soldier in the Middle Ages. In our study, we observed this type of ``unexpected creation'' happening with many of the examples, including a simple circle turning into a balloon (l), a centaur inspired by a sword and shield (e), and an energy blast derived from an unexpected action trigger of a simple line (h). While a few participants had a clear vision at the beginning, like body annotation for biology and anatomy education (d), overall, the participants appreciated the ability of improvisational and spontaneous exploration enabled by freehand drawing and interactive experiments.

\subsubsection*{\textbf{4) Animation Techniques}}
We also asked for user feedback for each animation feature. 
From the questionnaire responses, the participants positively responded to most of the features regarding the usefulness and understandability of the creation process for each feature: 
object binding (6.4/7, 6/7),
flipbook animation (5.55/7, 4.95/7),
action trigger (6.05/7, 5.45/7),
particle effects (5.25/7, 5.55/7),
motion trajectory (5.95/7, 6/7),
contour highlight (5.3/7, 6.45/7). 
We also counted the frequency of each animation technique used in the exploration study: object binding (14/20), flip-book animation (10/20), action trigger (9/20), particle effects (12/20), motion trajectory (7/20), contour lines (8/20). 
From our observations, most participants began by binding a simple shape and then gradually added more complex animations, such as emitting snowflakes, to refine their scenarios. Some participants chose not to use our contour highlight, particle effects, and motion trajectory features but instead used flip-book animations to achieve visually similar results by considering the depth effects.
From the questionnaire response and usage data, overall, the participants liked all six animation techniques.

The participants also provided comments on individual features.
\textit{``P15 on flip-book animation: This feature was particularly useful when paired with other features such as action-trigger. It is relatively simple to create with a few seconds of instruction.''} 
\textit{``P19 on action trigger: I really liked the action-trigger feature. I see this as being the most useful feature coupled with the frame drawing feature.''} 
\textit{``P4 on motion trajectory: cool feature, could be used in videos to make movements more fancy and interesting.''} 
\textit{``P17 on particle effects: Very creative feature for lasting animation like weather and environment effects.''}
\textit{``P12 on contour highlight: Easy to implement and useful to extract just the object you want to animate.''}


\subsubsection*{\textbf{5) Potential Use Cases}}
The participants found \system{} useful for social media (6.1/7) and storytelling for children (6.2/7).
All participants saw potential benefits of \system{} for classroom teachings (5.35/7).
\removed{In particular, participants felt that \system{} provided an ``P3: Easy and engaging educational videos and lab tutorials can be created using \system{} which will be easy for teachers to create and helpful for students. Interactive and engaging experience for the students to visually see and involve in creating the abstractions of concepts''.}
In particular, the participants felt that \system{} provided an easy and engaging way to create educational videos and lab tutorials, which can be helpful for both teachers and students (P3). They also expressed interest in using \system{} for a variety of other applications, including presentations (P8), sports analysis (P15), and safety and training demonstrations (P17).

\section{Expert Review}
\subsection{Method}
We also conducted an expert review to gain in-depth feedback. We recruited seven experts mainly from two groups - video creator experts and experts with a background in theatre. 
E1: full-time animation artist who has 18K followers and 1.8M upvotes on TikTok, 
E2: professional video creator who has 100K followers on BiliBili,
E3: professional video creator who has 97K followers on YouTube,
E4: researcher in AR and animation creation fields in a renowned company,
E5: theatre performer with 10 years experience, also a Computer Science Ph.D. student with interests in Mixed Reality
E6: theatre performance director/creative head with 25+ years of experience
E7: theatre director and actor with 5 years of theatre experience

We interviewed video creators to qualitatively compare our system with existing tools. While for experts with a background in theatre, we evaluate our system for improvisation, storytelling, and AR prototyping perspectives. We recruited theatrical performers and directors strongly interested in creating AR-based live performances via university mailing lists. We conducted a workshop to let them use our system to explore and prototype their AR-based performance so that they have a better understanding of the different features our tool offers and the workflow that our tool can support. Designers and technical developers also joined the workshop \added{and all  participants used the tool in pairs or groups}, thus ensuring an opportunity for collaboration among professionals from different backgrounds and with different skill sets. We were specifically interested in evaluating the qualitative measurements of the improvisational ability of our system. After the workshop, we conducted in-person interviews with each participant to reflect on their prototyping and ideation experiences and gain insights into how they can use our system for augmented live performances and storytelling. The interview lasted approximately one hour for each expert, and we provided \$20 CAD for their participation.

\subsection{Results}
All experts were excited by the potential of our tool. They found the tool to be intuitive (E1, E3,E4), useful (E3, E5, E7), playful (E1, E6), compelling (E2), general purpose (E1), and versatile (E3). Overall, the experts found the system enabled a creative, improvisational, and expressive range of freehand scribble animations, which they highlight in the strengths and limitations of our system.

\subsubsection*{\textbf{1) Comparison to Current Practices and Workflows: }}
Experts found the authoring workflow to be intuitive. They thought the tool could greatly reduce the time and cost of adding embedded animations. \textit{``E2: It takes considerable time to add effects in Adobe AfterEffects than this system's process of sketching while recording''.} E4, an HCI expert, pointed out that \system{} has a unique proposition of real-time creation and interaction. \textit{E4: ``Current tools and research projects are either made for the pre-production or post-production stage of content creation.''}. In addition, experts found that the identified six animation techniques were extensive and covered the most commonly seen scribble effects (E1-E4). Theatre experts saw a similarity between the director-performer relationship in theatre and the animator-actor dynamic between tool users (E5-E7). For example, just like the director, the animator directs the actor and makes decisions about the scene, props, and story. Overall, experts thought that our tool could complement their existing workflows.

\subsubsection*{\textbf{2) Encourages Discovery and Ideation: }}
Experts shared that \system{} could be used for ideation as you can rapidly test, discard and recreate ideas (E1, E3, E4). 
\textit{E3:``I can see it as an AR prototyping tool, not for the final product but a quick lo-fi mockup which enables idea generation''.} Experts also found that our system's real-time feedback helps ensure that their ideas are technically feasible. \textit{E5:``We spend time coming up with gestures and actions only to find they are not being detected. The ability to interact in real-time allows us to make these beautiful discoveries because now you have some specifics available, instead of making whatever we like now and thinking about the feasibility later''}
Other than ideas for augmented performances, experts also found that the tool could be useful in traditional theatre. \textit{E7:``You can also sketch props, costumes, and set designs and even how they interact with each other, which could be especially useful in rehearsals.''} 

\subsubsection*{\textbf{3) Improves Communication and Collaboration:}}
An important aspect that experts highlighted was that it could help improve communication and collaboration, specifically by reducing miscommunication between teams. \textit{E6:`` ... and he and I never got to the same visual language. So I think a shared language like this would be really helpful.''} As E6 and E7 had been directors, they remarked that it would benefit directors as they often need to communicate with multiple teams with different skills and technical languages.

\subsubsection*{\textbf{4) Lightweight Design: }}
Experts appreciated that \system{} is lightweight with a mobile AR web-based interface and requires minimal setup.
Video creators appreciated that the system was quick and did not require expertise. \textit{``E5: I could imagine the world would become more interesting with \system{} on a mobile phone. I could just go out to the streets and shoot a movie with animations and effects''}. The experts with a theatre background also appreciated the low-cost setup (E5-E7) as they pointed out that small theatre groups often have low budgets (E6, E7). \textit{ E6:``We don't want to bring in a ton of projectors or control lighting and other tedious and costly arrangements. So making  it available on mobile and a wider audience is precisely why it is so exciting!''}. Experts also appreciated the freehand sketched modality of embedded animations as it gives more agency over the visuals, adds personality, and makes the video look authentic. \textit{E3:``I have worked with some designers who want to add sketches to their videos as it gives a lot of personality to the video and makes it authentic.}


\subsubsection*{\textbf{Limitations and Future Suggestions}}
On the other hand, the experts also shared many limitations of \system{},  which were not identified in the usability study. For example, E2, E3, and E5 raised the concern that \system{} may not be suitable for those lacking sketching expertise.\textit{``E2: the prototype is only usable for artistically inclined users.''}. All the experts agreed that freehand drawing provides freedom and flexibility but requires the user to draw a new sketch every time they use it. Experts shared that the system needs a few modifications to replace video editing platforms (E1-E4). \textit{``E4: It does not have a video and audio track and other video editing functionalities''}. Also, although our system allows the user to erase the sketches in the drawing stage, the created effect cannot be removed once applied to the system. \textit{``E1: For a drawing system, an 'undo' button is essential.''} and \textit{``E2: I strongly request a fully functional 'undo' button.''}.

\subsubsection*{\textbf{Improvisational vs. Preparation-Based Authoring}}
Based on the rich insights we gained from the expert interviews, we reflected on the contrast between improvisational and preparation-based authoring systems. Improvisational authoring tools provide users with more flexibility during the creation process, allowing them to create and modify content in real time without the need for prior planning and preparation. This can be useful for quickly exploring different ideas and possibilities, particularly during the ideation phase. However, improvisational tools can result in lower production quality as a trade-off. Also, improvisational performances can be challenging to control and duplicate, resulting in undesired animation sequences.
In contrast, preparation-based tools give users more control over the final product but can be inflexible and time-consuming. Choosing between improvisational and preparation-based tools depends on the user's goals and needs. Future research could explore tools that allow users to switch between improvisational and preparation-based approaches or combine the best features of both in a single authoring tool.

%% file: 7-future-work.tex
\section{Limitations and Future Work}
On top of the feedback and suggestions from the usability study and expert interview, we also see many potential future work directions to expand our approach.




\subsubsection*{\textbf{3D Sketched Animation with HMDs}}
Our current implementation uses screen-based mobile AR, but integration with head-mounted displays (HMDs) like Microsoft Hololens or Meta Quest Pro will allow more blended and immersive experiences through mid-air drawing like \textit{Norman}~\cite{norman}.
However, this requires a more complex implementation, as it needs 3D object deformation, as opposed to a simple 2D line animation (e.g., \textit{Motion Amplifier}~\cite{kazi2016motion} vs. \textit{Layered 3D Animation}~\cite{ma2022layered}).
Future work should explore these 3D animations through either 3D deformation~\cite{ma2022layered}, simple frame-by-frame animations~\cite{norman}, or using 2D sketches on an invisible surface embedded in 3D space~\cite{kaimoto2022sketched}. 


\subsubsection*{\textbf{Exploring More Complex and Multi-Modal Action Triggers}}
Currently, the system only supports a simple image-based action trigger (e.g., the distance between two points), which limits complex animation behaviors for many use cases (e.g., storytelling, AR prototyping, etc.). 
If the system could support more complex and expressive action triggers, like the capability to define multi-modal triggers or a combination of multiple triggers, it would further expand possibilities. For example, if we could use a specific sound as a trigger, like a speech bubble animation when talking or experiences similar to \textit{Teachable Reality}~\cite{monteiro2023teachable}, \textit{AR Music Visualizers}~\cite{tanprasert2022ar}, and \textit{Music Bottles}~\cite{ishii2001bottles}). 


\subsubsection*{\textbf{Leveraging Real-time Parameters}}
In previous works of real-time responsive sketches like \textit{RealitySketch}~\cite{suzuki2020realitysketch}, \textit{Interactive Body-driven Graphics}~\cite{saquib2019interactive}, \textit{GesturAR}~\cite{wang2021gesturar}, or \textit{Reactile}~\cite{suzuki2018reactile}, the real-world parameters are often incorporated (e.g., an idea of manipulating sketches based on a distance between hands has been explored in \textit{GesturAR}~\cite{wang2021gesturar}).
In our prototype, we did not explore this aspect except for a simple action trigger to make our interaction and system simple, but the integration of real-time parameters could enhance the expressive animation.
For example, based on the distance between two hands, the user could change the scale or sketched objects, rotate the objects, or change the number of particles in dynamic and responsive ways. 

\subsubsection*{\textbf{Integration with Physics Simulations and Other Pre-defined Animations}}
The way to create responsive sketched animation can be situated in the spectrum between user-defined and pre-defined animations~\cite{suzuki2020realitysketch}.
Currently, our sketched animation focuses more on the user-defined aspect, in which the user can define most of the animations on demand.
However, the integration with pre-defined animations can enrich the expression. 
For example, the system could integrate physics simulations to make the sketched animations interact with the real world, similar to \textit{Sketched Reality}~\cite{kaimoto2022sketched}.
Another pre-defined animation like character motion or pre-programmed behavior like \textit{RakugakiAR}~\cite{rakugakiar} or \textit{ChalkTalk AR}~\cite{perlin2018chalktalk, perlin2018chalktalkvrar} is another possibility to broaden the expression.
Future work can investigate this aspect to create more complex and expressive animation.

%% file: 8-conclusion.tex
\section{Conclusion}
In this paper, we introduced \system{}, an augmented reality sketching tool to bring life to AR sketches based on real-world interactions. 
The design of our system is informed by the taxonomy and design space exploration of existing scribble animations. 
To support these animation techniques, this paper contributes a set of interaction techniques that enable real-time sketched animation without preparation or pre-defined configurations. 
Our user study confirms that our tool is easy to use yet enables expressive sketched animation.
We discussed limitations and future work to highlight broader challenges to enable more expressive AR sketched animation.